\documentclass[runningheads]{llncs}
\usepackage{amsmath}
\usepackage{graphicx}
\usepackage{array,xcolor,colortbl}

\usepackage{hyperref}

\newcommand{\zCOSMOS}{\textit{zCOSMOS}}
\newcommand{\SFR}{\textit{SFR}}
\newcommand{\SFRb}{\mbox{\textit{SFR}$^{1/12}$}}
\newcommand{\mass}{$M_\odot$}
\newcommand{\age}{$A$}
\newcommand{\RA}{$\alpha$}
\newcommand{\dec}{$\delta$}
\newcommand{\maga}{$M_V$}
\newcommand{\magb}{$-M_V$}
\newcommand{\LBT}{$t_L$}

\newcommand{\examples}{\url{https://www.lim.di.unimi.it/demo/zcosmos.php}}

\newcolumntype{C}{>{\centering\arraybackslash}p{0.15\textwidth}}
\newcolumntype{S}{>{\centering\arraybackslash}p{0.1\textwidth}}
\newcolumntype{B}{>{\centering\arraybackslash}p{0.2\textwidth}}

\begin{document}
\title{A Sonification of the zCOSMOS Galaxy Dataset}
%

\author{Sandro Bardelli\inst{1}\orcidID{0000-0002-8900-0298} \and
Claudia Ferretti\inst{2} \and
Giorgio Presti\inst{3}\orcidID{0000-0001-7643-9915} \and
Maurizio Rinaldi\inst{2}}
\authorrunning{S. Bardelli et al.}
%
\institute{Osservatorio di Astrofisica e Scienza dello Spazio\\
INAF -- National Institute for Astrophysics\\
\and
Sound artist, sensorialist, expert of perception processes
\and
LIM -- Laboratory of Music Informatics\\
Department of Computer Science ``Giovanni Degli Antoni''\\
University of Milan
\and
electornic music performer
}
\maketitle              
\begin{abstract}
This paper proposes a sonification for \zCOSMOS, an astronomical dataset that contains information about 20,000 galaxies. The goals of such an initiative are multiple: providing a sound-based description of the dataset in order to make hidden features emerge, hybridizing science with art in a cross-domain framework, and treating scientific data as cultural heritage to be preserved and enhanced, thus breaking down the barriers between scientists and the general audience. In the paper, both technical and artistic aspects of the sonification will be addressed. Finally, some relevant excerpts from the resulting sonification will be presented and discussed.

\end{abstract}

\section{Introduction}

\textit{Sonification} is the transformation of data into acoustic signals, namely a way to represent data values and relations as perceivable non-verbal sounds, with the aim to facilitate their communication and interpretation \cite{kramer1999sonification}. Like data visualization provides meaning via images, sonification conveys meaning via sound.

As discussed in \cite{barrass1999using}, non-verbal sounds can represent numerical data and provide support for information processing activities of many different kinds. 

A first scenario is the possibility to receive information while keeping other sensory channels unoccupied, as required in medical environments, process monitoring, driving, etc. Common experiences in everyday life range from the sounds naturally produced by physical phenomena and automatically associated with specific events (e.g., a whistling kettle) to sound-augmented objects (e.g., a Geiger counter). This approach is explicitly used in sensory-substitution systems, like orientation and navigation applications for blind or visually impaired (BVI) people \cite{aabbglmp2018sonificationof}.

Sonification techniques prove to be useful also when the data to represent are complex and have multiple dimensions to track \cite{ben2001natura}. In fact, music and sound present multidimensional features (e.g., pitch, intensity, timbre, spatialization, etc.), and these dimensions can be simultaneously employed to provide understandable representations of complex phenomena. An effective design of sonification can draw, e.g., on musicality, musical acoustics, sound synthesis and human perceptual capacities \cite{bregman1994auditory}. Listening to data can open new scientific frontiers, thanks to the human ability to parse sound for patterns and meaning. This approach can make non-trivial structures emerge and help to unveil hidden patterns \cite{cooke2017exploring}. Social processes \cite{lenzi2020intentionality}, natural events \cite{arai2012sonification}, and physical observations \cite{dubus2013systematic} are only a few examples of applicability fields. 

Finally, sonification can be applied to those scenarios where a set of data is only the base to build an experience with artistic goals. In this context, it is worth mentioning also the concept of \textit{musification}, namely the representation of data through music. The resulting musical structures can take advantage of higher-level features, such as polyphony or harmony, in order to engage the listener. The relationships of sonification to music and sound art have been explored in \cite{gresham2012relationships}. After providing these definitions, we can affirm that data can be visualized by means of graphics, sonified by means of sound, and musified by means of music.

Sonifications can be enjoyed as scientific inquiries, aesthetic experiences, or both. The idea of bridging the gap between art and science in the context of scientific dissemination and ``edutainment'' initiatives has been explored in a number of works. An interesting point of view is reported in \cite{ballora2014wow}: according to the author, sonification, as opposed to visualisation, is still an under-utilised element of the ``wow'' factor of science.

Proposing a sonification initiative during an exhibition or another public event can also add value to the dataset itself. First, a sound-based multimedia installation can be an engaging way to make a non-expert audience enjoy scientific subjects; in this sense, the experience can be enhanced through suitable support materials (e.g., wall-mounted panels), the stimulation of other sensory channels (e.g., a video installation), and real-time interaction with the audience (e.g., through motion detectors and ambient-light sensors). Moreover, such an initiative can play a cultural role by raising scientific data and achievements to the rank of cultural heritage to be preserved and exploited. Examples are reported in \cite{avanzo2010data}, \cite{dunn1999life}, and \cite{polli2012soundscape}. 

In the context of sensory substitution techniques, sonification can make scientific data accessible to specific categories of users, e.g.\ BVI people, with an important impact on their education, too \cite{laconsay2020visualization,pereira2013sonified,reynaga2020strategies}.

For the sake of completeness, it is worth underlining that the legitimacy of sonification as a scientific method of data display is being debated by scholars and experts, as discussed in \cite{supper2012search}. According to \cite{neuhoff2019sonification}, widespread adoption of sonification to display complex data has largely failed to materialize, and many of the challenges to successful sonification identified in the past are still persisting. Nevertheless, since the goal of the initiative described below is dissemination, even if scientifically accurate, our proposal does not fall within the scope of problematic applications.

The rest of the paper is structured as follows: in Section \ref{sec:astronomical_data} we will review some background work about astronomical data sonification, in Section \ref{sec:zcosmos} we will present the \zCOSMOS\ galaxy dataset, in Section \ref{sec:strategy} we will describe our sonification strategy, in Section \ref{sec:discussion} we will discuss the achieved results, and, finally, in Section \ref{sec:conclusions} we will draw some conclusions.

\section{Sonification of Astronomical Data}
\label{sec:astronomical_data}

The idea of sonifying astronomical data is not original at all. A forerunner of such an approach is the \textit{photophone}, a device invented by Alexander Graham Bell that used light modulation, caught by means of photosensors, in order to transmit audio signals to a distant station \cite{bell1880photophone}. In a letter written in 1880, the inventor showed excitement about the possibility to ``hear a ray of the sun laugh and cough and sing''. He was intrigued by the idea of applying such a technology to study the spectra of stars and sunspots by listening to the sounds produced by the \textit{photophone} receiver \cite{brown2012listening}.

Many years later, space agencies promoted sonification as a mean to explore astronomical data. For example, NASA created a Java-based software tool called \textit{xSonify} \cite{candey2006xsonify,diaz2011sonification} aiming to encourage investigation in the field of space physics through sonification. Another initiative by NASA, dating back to 2020, is a project aiming to sonify the center of the Milky Way. Users can listen to data from this region captured by Chandra X-ray Observatory, Hubble Space Telescope, and Spitzer Space Telescope. Such data can be enjoyed either as solos or together, as an ensemble in which each telescope plays a different instrument. 

Lunn and Hunt \cite {hud15922} described case studies of astronomy-based sonification, specifically addressing the sonification of radio-astronomy data as part of the Search for Extra-Terrestrial Intelligence (SETI).

Hadhazy \cite{hadhazy2014heavenly} mentioned and exemplified some interesting musical compositions based on astronomical data. Among them, ``Deep-Space Sonata'' converted the gamma-ray burst GRB 080916C, one of the most powerful explosions recorded in the Universe, to audible sound. In this sonification, the number of notes played represents the gamma rays received by the Fermi Gamma-ray Space Telescope, while the accompanying sounds correspond to the probability of the rays emanating from the burst itself (lowest-likelihood rays are played as a harp, medium by a cello and highest-probability by a piano). 

Another experience reported in \cite{hadhazy2014heavenly} is ``Sunny Anthem'', based on the data of charged atoms within the solar wind from 1998 to 2010 recorded by a spectrometer onboard NASA's Advanced Composition Explorer spacecraft. 

Many other sonification experiments starting from astronomical data could be mentioned. For instance, ``Jovian Notes'' is a sonification captured by Voyager 1's plasma wave instrument as the spacecraft crossed the bow shock at the edge of Jupiter's magnetosphere. Another initiative was based on the data gathered by a scientific device onboard the Lunar Reconnaissance Orbiter concerning radiation spewed by the sun as it floods the vicinity of the moon; the radiation intensity, converted into musical sounds, can be considered as a sort of live lunar music. Other relevant experiences are discussed in \cite{2019aas}.

This kind of initiatives is also a way for stimulating interest when teaching astronomy. An example of auditory model of the solar system integrated within a planetarium show is described in \cite{tomlinson2017solar}. Moreover, Ballesteros and Luque \cite{ballesteros2008using} reported a successful application to the educational field, originally conceived for a science-dissemination radio program called ``The sounds of science'', heard on Radio Nacional de Espana (RNE), but easily reproducible also in a classroom environment. Finally, it is worth underlining that a sound-based approach can be particularly effective for BVI students \cite{ferguson2016bell3d}.

Concerning accessibility and inclusion for people who cannot fully enjoy experiences where visual media is the primary communication mechanism, it is worth mentioning also initiatives dealing with astronomic data but relying on senses other than hearing. An example is the creation of three-dimensional printed data sets, as in the case of tactile 3D models from NASA's Chandra X-ray Observatory \cite{arcand2019touching}. Also the sense of smell can be used to design astronomy-oriented programs, as reported in \cite{wenz2018scents}. Finally, the multi-sensorial project called ``The Hands-On Universe Project'' employs edible models and common food to convey complex concepts in cosmology and astrophysics \cite{trotta2018hands}.

Going back to sonification of astronomical data, there are also initiatives coupling dissemination purposes with artistic goals. For example, Ballora \cite{ballora2014sonification} applied sonification techniques for astronomical data in the framework of a film-making process, so as to create the soundtrack for a short movie titled ``Rhythms of the Universe''. In \cite{ballora2014sonification} he described how the sonifications had been obtained from datasets describing pulsars, the planetary orbits, gravitational waves, nodal patterns in the sun’s surface, solar winds, extragalactic background light, and cosmic microwave background radiation.

\section{The \zCOSMOS\ Dataset}
\label{sec:zcosmos}

\begin{figure}[ht!]
	\centering
	\includegraphics[width = 0.60\columnwidth]{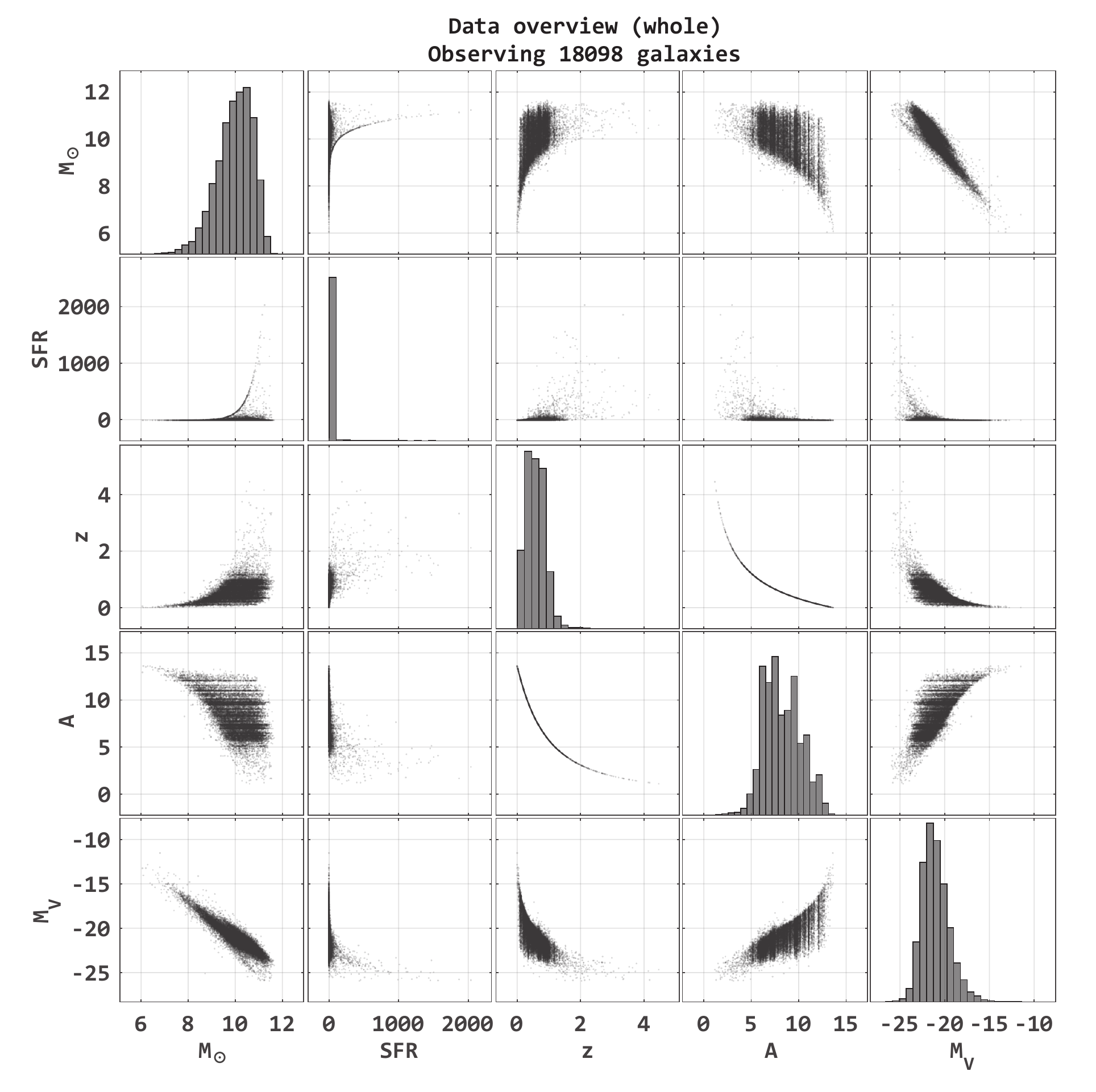}
	\caption{Overview of the original variables. Variables distribution histograms are shown on the main diagonal; other positions display a dispersion graph of each pair of variables.}
	\label{fig:plotmat1}
	
	\vspace{1cm}
	
    \includegraphics[width = 0.60\columnwidth]{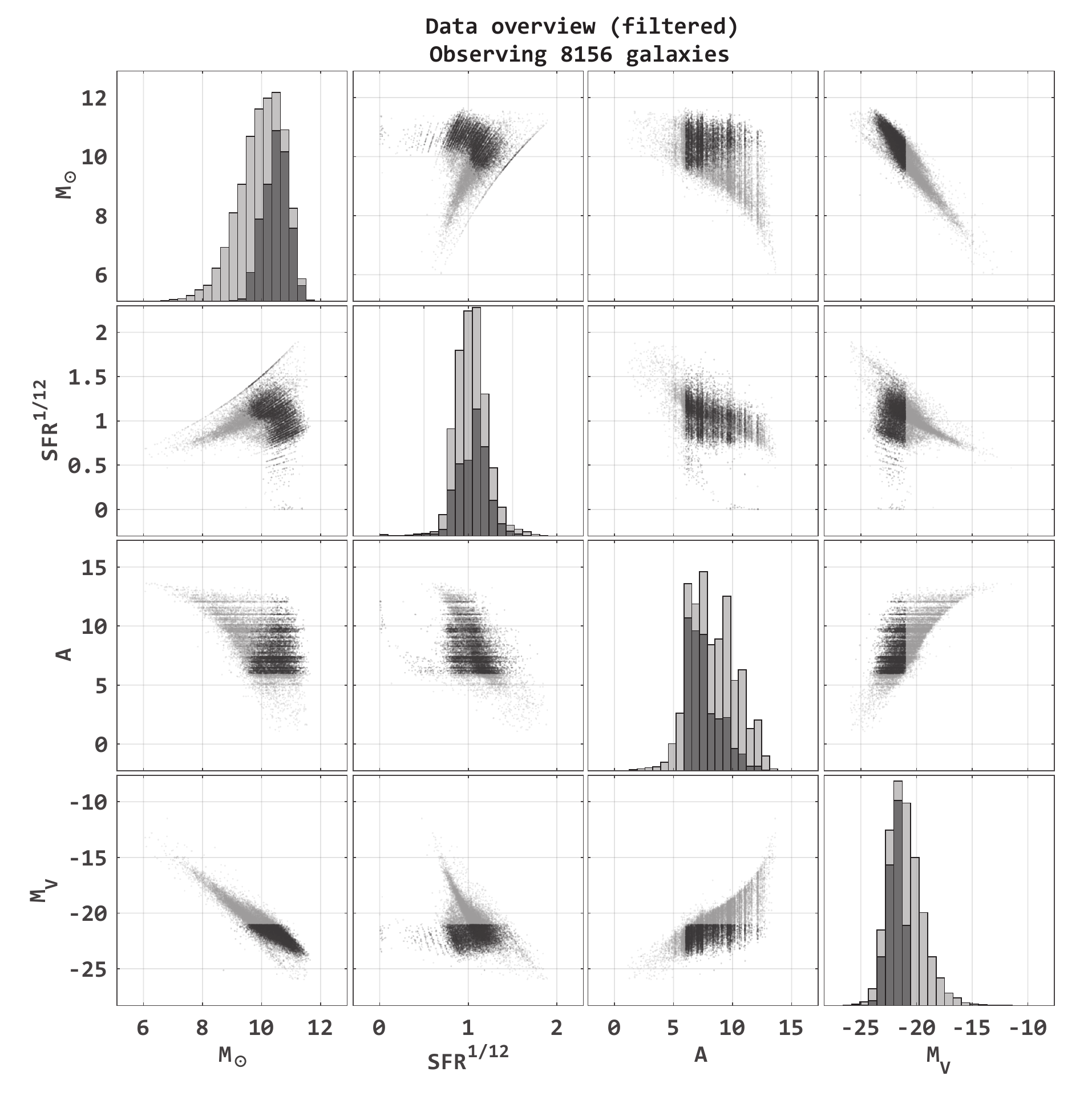}
	\caption{Overview of the re-scaled and filtered variables.  Galaxies considered for sonification are shown in dark color, while discarded galaxies are shown in light gray.}
	\label{fig:plotmat2}
\end{figure}

The \zCOSMOS\ dataset \cite{knobel2012zcosmos,lilly2009zcosmos,lilly2007zcosmos} 
is the spectroscopic follow-up of the wider \textit{Cosmic Evolution Survey} (\textit{COSMOS}) \cite{scoville2007}, a coordinated international effort to study the galaxy evolution in various wavelength. \zCOSMOS\ obtained spectra for 18,143 galaxies at the apparent magnitude $I_{AB}<22.5$ at the VLT-ESO telescope (Chile). Such a dataset describes the evolution of a relatively small portion of the Universe in the last 10 million years.

With the use of these high-quality spectra, it was possible to derive a number of physical quantities of the galaxies.
Among the variables provided in the dataset, we selected:

\begin{itemize}
	\item Stellar Mass \mass, a value describing how many stars 	are formed in a galaxy and, therefore, a proxy for the galaxy history. It is the sum of all the star masses;
	\item Star Formation Rate \SFR, i.e.\ the total mass of stars formed per year, which reflects how active a galaxy is at the moment of the observation;
	\item Redshift $z$, namely the measure of the recession velocity of the galaxy as a consequence of the expansion of the Universe. Due to the Hubble's law, the higher the redshift (measured as the shift of spectral lines toward the red part of the spectrum), the higher the 
	galaxy distance;
	\item Age of the Universe \age\ at the galaxy position, measured in billions of years. Objects that are close to us are observed with little time delay due to the finiteness of the speed of light; therefore, the value of \age\ for this galaxy is similar to the actual age of the Universe (i.e. about $13.8$ billion years). Conversely, distant objects are seen with a remarkable delay, thus, in our observations, \age\ at that distance is lower. In other terms, even if we are observing objects at a given moment, depending on their distance we are observing a younger or a later stage of the Universe. \age\ is linked to a measure called \textit{lookback time}, hereafter referred to as \LBT, obtained by subtracting \age\ from the age of the Universe: \LBT\ $=13.8-A$. The lookback time of a galaxy represents how much time we are looking back to obtain its current observation;
	\item Absolute magnitude \maga, i.e.\ the absolute luminosity of the observed galaxy (related to the intrinsic luminosity). Please note that lower values of \maga\ imply brighter galaxies;
	\item Position, in terms of right ascension \RA\ and declination \dec. These are the coordinates of the galaxy over the celestial sphere in the equatorial coordinate system.
\end{itemize}

The number of galaxies under exam has been reduced to 18,098, since the ones presenting $M_V < -26$ or $M_\odot \le 6$ could come from measurement errors or insufficient quality of data.

Prior to the choice of a sonification strategy, we conducted a simple statistical inspection of the dataset. In fact, a sonification is expressive when changes in sound significantly reflect those in data. For this to happen, it is important that such data take values with a sufficient resolution and within a perceptible range. Furthermore, in case of a strong correlation between some variables, we had to understand if the reason was trivial, e.g.\ the calculation of one datum from another, or rather it was an unexpected feature of the phenomenon. In the former scenario, the sonification should avoid redundancy, while, in the latter, it should remark these characteristics.

The top-left to bottom-right diagonal of Figure \ref{fig:plotmat1} shows the statistical distribution of variables. Its analysis allows to infer the range of variation of each variable. The other areas of Figure \ref{fig:plotmat1} depict the dispersion graphs of each pair of variables, so as to remark their mutual dependence. Three problematic issues mainly emerge: 

\begin{enumerate}
	\item Almost all values for \SFR\ are condensed in the leftmost area;
	\item Variables $z$ and \age\ are, not surprisingly, one function of the other, since $z$ is a proxy for \LBT\ in standard cosmological models;
	\item Due to an observational bias known as Malmquist bias \cite{malmquist1922some}, at large distances the sample loses faint galaxies;
\end{enumerate}

In Section \ref{subsec:design} we will describe the design choices we implemented in order to solve these problems.

\section{Sonification Strategy}
\label{sec:strategy}

Sonification can be seen as the junction point between the artistic use of science and the scientific use of art, thus combining the separate viewpoints of the artist and the scientist. Coherently, the project described in this paper has been designed and implemented by a working group made of scientists, technicians, and artists.

The software tools used in the whole process include: MATLAB for data inspection and preprocessing, Supercollider to parse the CSV exported by MATLAB and perform real-time sound synthesis, Ableton Live to record the distinct audio tracks generated by Supercollider, and, finally, Steinberg Cubase for post-production.

As better detailed below, the proposed sonification is based upon three main layers: 

\begin{itemize}
    \item Galaxies, sonified through a dense stream of events, each modulated independently, thus generating a synthetic sound texture (Section \ref{sec:galaxies});
    \item Statistics, producing a very simple, continuously modulated, synthetic drone sound (Section \ref{sec:stat});
    \item Outliers, causing a rare occurrence of events, each modulated independently, thus generating complex sound icons (Section \ref{sec:outliers}).
\end{itemize}

Sonification examples of the listed items and portions of the final outcome can be found at the following URL: \examples.

\subsection{Data Pre-Processing}
\label{subsec:preprocessing}

In Section \ref{sec:zcosmos} we listed three issues emerging from an \textit{a-priori} statistical inspection of the dataset. These problems have been addressed in a pre-processing phase.

In order to solve the first issue, namely an excessive clustering of \SFR\ values in the leftmost range, an early idea was to consider \mbox{$\log$(\SFR)} instead of \SFR, but this would have involved an excessive flattening of the values to the right. We therefore decided to transform the data in a monotonic way by extracting the twelfth root of \SFR, namely \SFRb, which showed a more balanced distribution (see Figure \ref{fig:plotmat2}).

The second issue, namely the correlation between $z$ and \age, was solved by ignoring $z$, whose flattening to the left was higher than the one of \age. This is the reason why $z$ is not present in Figure \ref{fig:plotmat2}.

The \zCOSMOS\ survey is limited in apparent luminosity and therefore at higher distances it  observed only the brightest galaxies.
In order to avoid this observational bias and to have an homogeneous sample,  we decided to limit the time span and the brightness range: galaxies with \mbox{\age $< 6$} million years and $M_V > -21$ have been excluded. In this way, the number of sonified galaxies has been drastically cut, from $18,098$ to $8,156$. Even if the sonification of the full set of galaxies could have been desirable from multiple points of view, we chose not to represent misleading data, such as the false correlation between magnitude and age. The remaining galaxies are visible in dark color in Figure \ref{fig:plotmat2}. 

The reduction in the number of represented galaxies pushed us to reflect on the opportunity to provide an acoustic feedback to uncertainty in observations. The mechanism employed to achieve this goal will be described in Section \ref{sec:stat}.

Concerning \age, as mentioned before, such a parameter has been converted into \LBT, since the latter will better suit the design described in Section \ref{sec:galaxies}.

Another adjustment has involved \maga, that has been inverted in order to handle values in a more intuitive way. This parameter, indicated as \magb\ from now on, takes low values for dim objects and high values for bright ones.

\begin{figure}[ht!]
	\centering
	\includegraphics[width = 0.68\columnwidth]{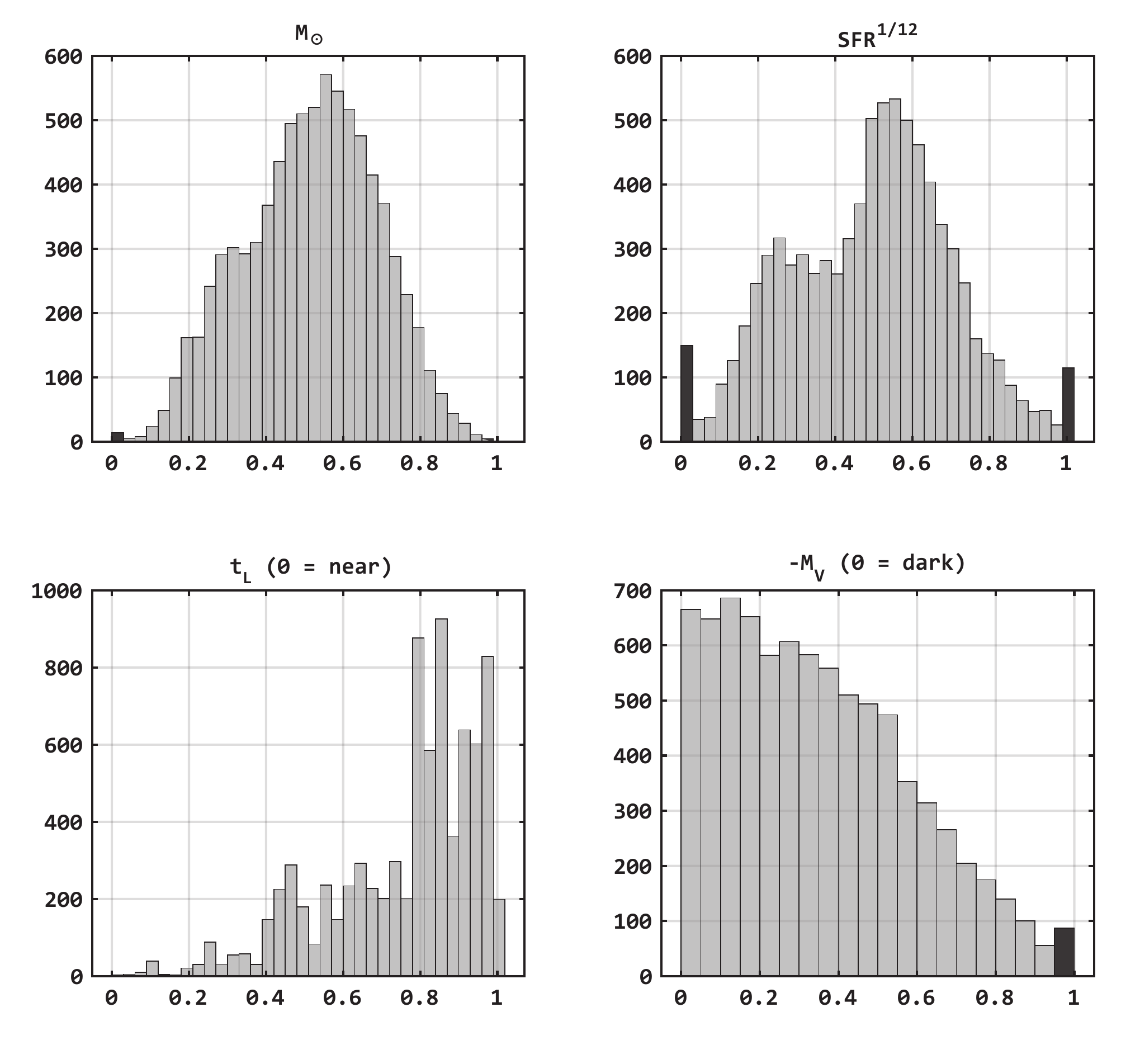}
	\caption{Distributions of the normalized variables that are actually sonified. The uniformly distributed variable \RA\ is not shown. Outliers assigned to auditory icons are highlighted through dark histogram bars.}
	\label{fig:finalplotmat}
	
	\vspace{0.4cm}
	
	\includegraphics[width = 0.68\columnwidth]{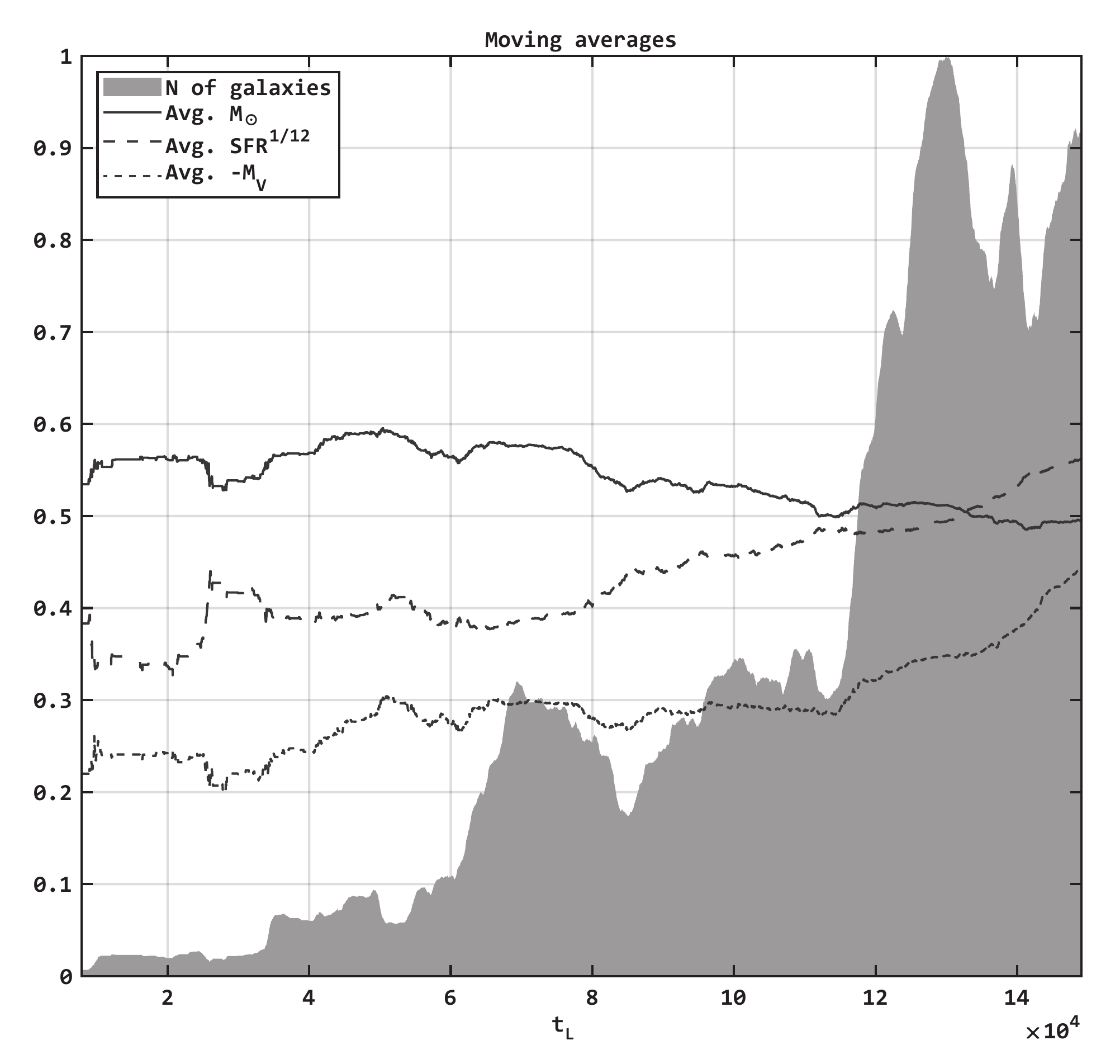}
	\caption{Moving average of principal features against lookback time.}
	\label{fig:stats}
\end{figure}

Features have been normalized in the interval $[0,1]$ so as to provide a simple interface to the modular sonification engine. In order to avoid very small and very big outliers that would shrink most values into a middle range, some clipping has been introduced. Clipped values have been marked, so as to be properly treated in the sonification. Please refer to Table~\ref{tab:clip} and Figure~\ref{fig:finalplotmat} for further details.

\begin{table}[tb]
\caption{Lower and upper thresholds and number of outliers. Dim galaxies, presenting high \maga, have not been marked.}
\centering
\begin{tabular}{CBBBB}
\hline
\textbf{Variable} & \textbf{Lower threshold} & \textbf{No. of lower outliers} & \textbf{Upper threshold} & \textbf{No. of upper outliers} \\ \hline
\mass             & 9.25         & 7               & 11.58        & 1               \\
\SFR              & 0.1          & 138             & 76           & 102             \\
\maga             & -24.19       & 9               & -            & -               \\ \hline
\end{tabular}
\label{tab:clip}
\end{table}

As the final step, a moving average has been calculated for normalized \mass, \SFRb, and \magb\ to provide explicit information about the trend of these variables along the temporal axis. The results thus obtained are visible in Figure~\ref{fig:stats}.

\subsection{Design Choices}
\label{subsec:design}

The sonification is completely procedural and parametric, so it can be easily modified to respond to different requirements and use cases. In particular, modularity has been exploited to explore the space of sound-synthesis parameters, overall duration, and feature associations, with the aim of matching the sonification goals with the desired aesthetics.

In all design phases we followed the principle of \textit{ecological metaphors}, thus trying to make the sonification coherent with users' real-world sensory and cognitive experience~\cite{brunswik1953ecological}. This approach implies that variations of auditory dimensions are consistent with those of physical parameters; for example, position values may be mapped onto left/right sound panning. The use of ecological metaphors should improve intuitiveness and learnability.

One of the goals was to keep the technical setup required to play the sonification as simple as possible, so as to make it easily reproducible in a wide range of contests. For this reason, the final outcome was a standard stereo file. Sound spatialization through an array of loudspeakers or a binaural approach would have extended the possibilities connected to ecological metaphors, but it would have prevented the performance in a great number of environments not adequately equipped.

\subsection{Sonification of Galaxies}
\label{sec:galaxies}

Each galaxy is sonified as a single and short sound event, occurring at a time which is proportional to its lookback time. When the event density is very low, single galaxies can be easily spotted and compared, while, in case of very dense and crowded sections, the overlapping of many events generates a complex texture which is more informative about the overall trend.

Each sound event is generated by $3$ distinct sinusoidal oscillators, called $O_{1-3}$, presenting an exponential-decay envelope. Each oscillator can be controlled in terms of pitch, level, decay time, stereophonic position, and frequency modulation. $O_1$, $O_2$, and $O_3$ are the master, the harmonic, the sub-harmonic oscillator respectively. After computing the parameters for $O_1$, those for the other oscillators follow according to the schema shown in Table~\ref{tab:osc}. The rationale is to have a fundamental frequency generated by $O_1$ which is louder and lasts longer than the harmonic and sub-harmonic sounds generated by $O_{2,3}$.

\begin{table}[tb]
\caption{Relationship between the parameters of the oscillators.}
\centering
\begin{tabular}{CCCCCC}
\hline
\textbf{Oscillator} & \textbf{Pitch} & \textbf{Level} & \textbf{Release} & \textbf{Position} & \textbf{FM} \\ \hline
\textbf{$O_1$} & $f_0$     & $a$     & $r$    & $p$ & None \\
\textbf{$O_2$} & $2f_0$    & $0.25a$ & $0.5r$ & $p$ & None \\
\textbf{$O_3$} & $0.75f_0$ & $0.6a$  & $0.6r$ & $p$ & Yes  \\ \hline
\end{tabular}
\label{tab:osc}
\end{table}

As it regards $O_1$, its parameters are modulated according to the following data bindings (a detailed description of modulation ranges is presented in Table~\ref{tab:galpar}).

The volume is determined by the absolute magnitude \magb\ of the galaxy under exam by exploiting an analogy with vision: brighter galaxies are represented with louder sounds, while dim galaxies (harder to see) are represented with softer sounds (harder to hear). The resulting dynamic range is about $-24$dB, which is sufficient to discriminate between bright and dim galaxies without making the latter inaudible.

Since the sonification has been conceived to be reproduced through a stereophonic speakers layout, it was natural to bind galaxy right ascension with the sound position in the stereophonic space. The resulting representation of spatial information is magnified, since original right ascension of the galaxies is included in about $1$ degree of the sky, while the stereophonic field can reach $180$ degrees, depending on the installation conditions. Declination \dec\ could have been treated in a similar way, thanks to quadraphonic listening environments, but we decided to privilege a simpler setup, as explained in Section \ref{subsec:design}.

The frequency of $O_1$, namely $f_0$, is inversely proportional to \mass. This binding has been chosen since lower pitches are generally associated with heavier and bigger sources, while high-pitched sounds easily recall smaller sources. In order to produce well-sounding events, many sonifications (including the ones mentioned in Section \ref{sec:astronomical_data}) usually map values onto notes of the equal-tempered scale or consonant frequencies. Conversely, we decided to let the frequency binding be continuous; in this way, the presence of beatings as opposed to the perception of distinct sounds lets the listener clearly perceive when two galaxies are similar (beatings) or different (distinct sounds) in terms of \mass. As an aesthetic consideration, the adoption of a musical scale would have produce a sonification more pleasant in the short term, but more boring on the long run. Another potential problem was the possibility to introduce a phenomenon of data misinterpretation in case of peculiar musical structures  (e.g., consonant chords, cadences, etc.), which are strongly rooted in the tonal-harmony perception of music, but have no particular meaning in the sonification.

Star formation rate \SFR\ is linked to the parameters of the frequency modulation of $O_3$. In order to give the idea of very active galaxies for high values of \SFRb\ and more relaxed galaxies for low values of \SFRb, we carefully tuned the sinusoidal modulator frequency $f_m$. This parameter runs below the audio rate (i.e. $f_m<20$ Hz) for low star formation rate, thus producing a tremolo-like effect, while high values for \SFRb\ produce a more distinctive and frantic modulation. For the same reason, the frequency deviation $d$, a measure of the frequency modulation amount \cite{dodge1985computer}, is modified proportionally to \SFRb, too.

Finally, the amplitude envelope exponentially decays with a factor proportional to \SFRb, so that galaxies with high \SFRb\ present a longer tail, while low values of \SFRb\ cause a quicker decay.

Table~\ref{tab:galpar} provides a detailed view of the parameters.

\begin{table}[tb]
\caption{Sonification parameters for a single galaxy.}
\centering
\begin{tabular}{CCSSCC}
\hline
\textbf{Variable}   & \textbf{Parameter} & \textbf{0} & \textbf{1} & \textbf{Unit}       &\textbf{Type}  \\ \hline
\LBT         & Time        & 0        & 1500      & s         & Linear      \\
\magb      & Level       & -34      & -10       & dB\textsubscript{fs} & Linear \\
\RA    & Position    & Left     & Right   & Pan            & Linear       \\
\mass       & $f_0$       & 7000     & 400     & Hz             & Exponential  \\
\SFRb       & $f_m$       & 2.88     & 252     & Hz             & Exponential  \\
\SFRb       & $d$  & 12       & 1050    & Hz             & Exponential  \\
\SFRb       & Release     & 0.3      & 9.6     & s        & Linear\\ \hline
\end{tabular}
\label{tab:galpar}
\end{table}

\subsection{Sonification of Statistics}
\label{sec:stat}

Statistics include the average of \mass, \SFRb, and \magb\ computed within a moving window across lookback time. These are continuous signals controlling the frequency $f_{1-3}$ of 3 distinct resonant bandpass filters, each one filtering white noise, with different pan values. Filters frequency are set by multiplying \mass, \SFRb, and \magb\ by $200$, $1000$, and $2000$ respectively. The result is a drone sound, a non-tempered chord which is consonant only under favorable circumstances.

The quality $Q$ of the filters is very high at the beginning of the sonification, thus producing well defined pitches; it linearly decreases in time, so as to produce band-limited noise at the end of the sonification. The idea is to suggest an increase in data variability and uncertainty of the observations as long as more distant time and space is under exam. Please note that the reciprocal of $Q$ is the actual modulated parameter.

\begin{figure}[tb]
	\centering
	\includegraphics[width = 0.8\columnwidth]{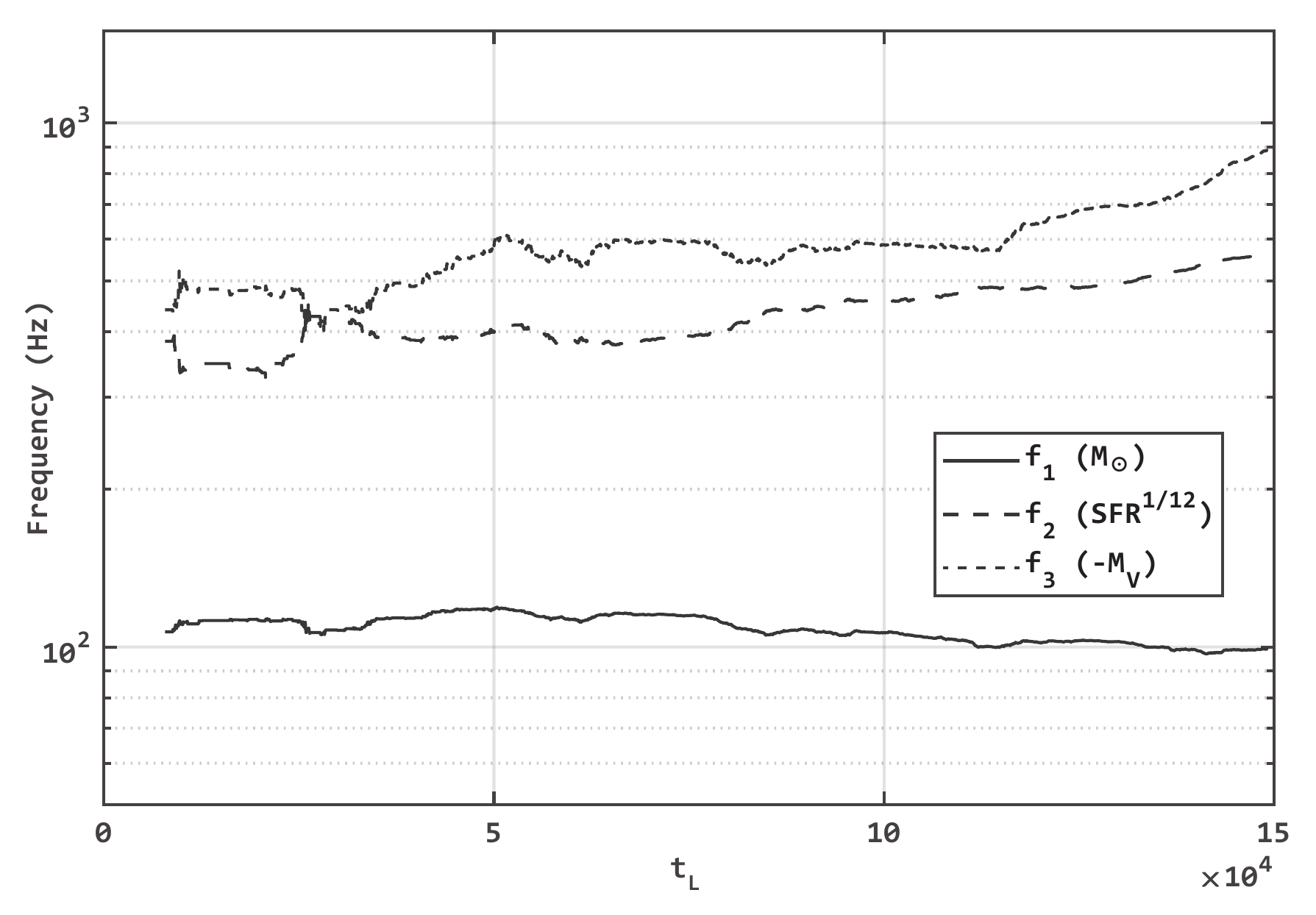}
	\caption{Actual frequencies played by statistics-controlled filters.}
	\label{fig:ft}
\end{figure}

Exact ranges are shown in Table~\ref{tab:statpar}, and the actual frequencies over time are visible in Figure~\ref{fig:ft}.

\begin{table}[tb]
\caption{Statistics sonification parameters.}
\centering
\begin{tabular}{CCSSSS}
\hline
\textbf{Variable}        & \textbf{Parameter} & \textbf{Min} & \textbf{Max} & \textbf{Unit} & \textbf{Pan}    \\ \hline
Avg. \mass     & $f_1$              & 97           & 120          & Hz            & Center   \\
Avg. \magb    & $f_2$              & 327          & 562          & Hz            & Left \\
Avg.\SFRb     & $f_3$              & 401          & 887          & Hz            & Right  \\
\LBT   & $Q_{1-3}$          & 0.0001       & 0.2          & $Q^{-1}$      & -      \\ \hline
\end{tabular}
\label{tab:statpar}
\end{table}

\subsection{Sonification of Outliers}
\label{sec:outliers}

Outliers are galaxies whose values are out of range for at least one variable\footnote{Actually, no galaxies were marked as outliers for more than one variable.} and are sonified by means of auditory icons, modulated (when possible) with the same principles of single-galaxies modulations.

The icons have been carefully crafted using sound design principles coherent with other sonification-design choices. The goal is to make outliers emerge from the overall sonification, but linking their perceptibility to the frequency of their occurrence: uncommon events have to stand out with respect to more common outliers. Please refer to Table~\ref{tab:clip} for the number of outliers of each kind.

The icon for the biggest galaxy is a low-pitched percussive sound, while smaller galaxies are associated with high-pitched bells; both approaches rely on the original binding, but provide more emphasis on their outlier nature. Such sounds are modulated in position and intensity, according to \RA\ and \magb.

Similarly, high and low \SFRb\ are represented by fast and slow pulsing rumbles, respectively. These are generated through filtered noise, and modulated in pitch, level, and position according to the original bindings, and release and non-linear distortion according to \mass\ and \SFRb\ respectively.

Finally, very bright galaxies are represented through sound glitches, so as to suggest a saturation effect for the sensors, modulated in pitch and position only.

\subsection{Post-Production and Final Outcome}

The post-production phase consisted in fine-tuning the level balance between galaxies, outliers, and statistics. Some reverberation and equalization was added in order to improve the aesthetic result.

In particular, more reverberation has been added to background drones respect to foreground sounds in order to create a sense of depth.

Since background drone sounds are modulated quite slowly, they were slightly processed with granular synthesis based effects, so to avoid adaptation effects and to render them more appealing, by providing a more ruffled texture without compromising the intelligibility of the played frequencies.

The final outcome was a 25 minutes long sonification, whose musical meaning will be discussed in the next section.

\section{Discussion}
\label{sec:discussion}

The proposed sonification can be formally framed according to the taxonomy outlined in \cite{ludovico2016sonification} and called the \textit{sonification space}. In that framework, any sonification is described as a set of \textit{data bindings} (i.e. the single mappings between data and sound features) and one \textit{main feature} (i.e. the overall sonic outcome). Both data bindings and the main feature can be placed over a two-dimensional plane characterized by the following axes: time granularity (continuous/regular/asynchronous), and level of abstraction (direct representation of data vs.\ symbolic representation). The areas that can be found are: 

\begin{itemize}
\item \textit{Soundscapes}, i.e.\ holistic descriptions of a system by means of continuously modulated symbolic sounds;
\item \textit{Symbolic samples}, i.e.\ asynchronous symbolic signaling such as auditory icons;
\item \textit{Feature modulations}, i.e.\ continuous or regular sonic plotting;
\item \textit{Sound events}, i.e.\ asynchronous and interactive sonic plotting.
\end{itemize}

\begin{figure}[tb]
	\centering
	\includegraphics[width = 0.8\columnwidth]{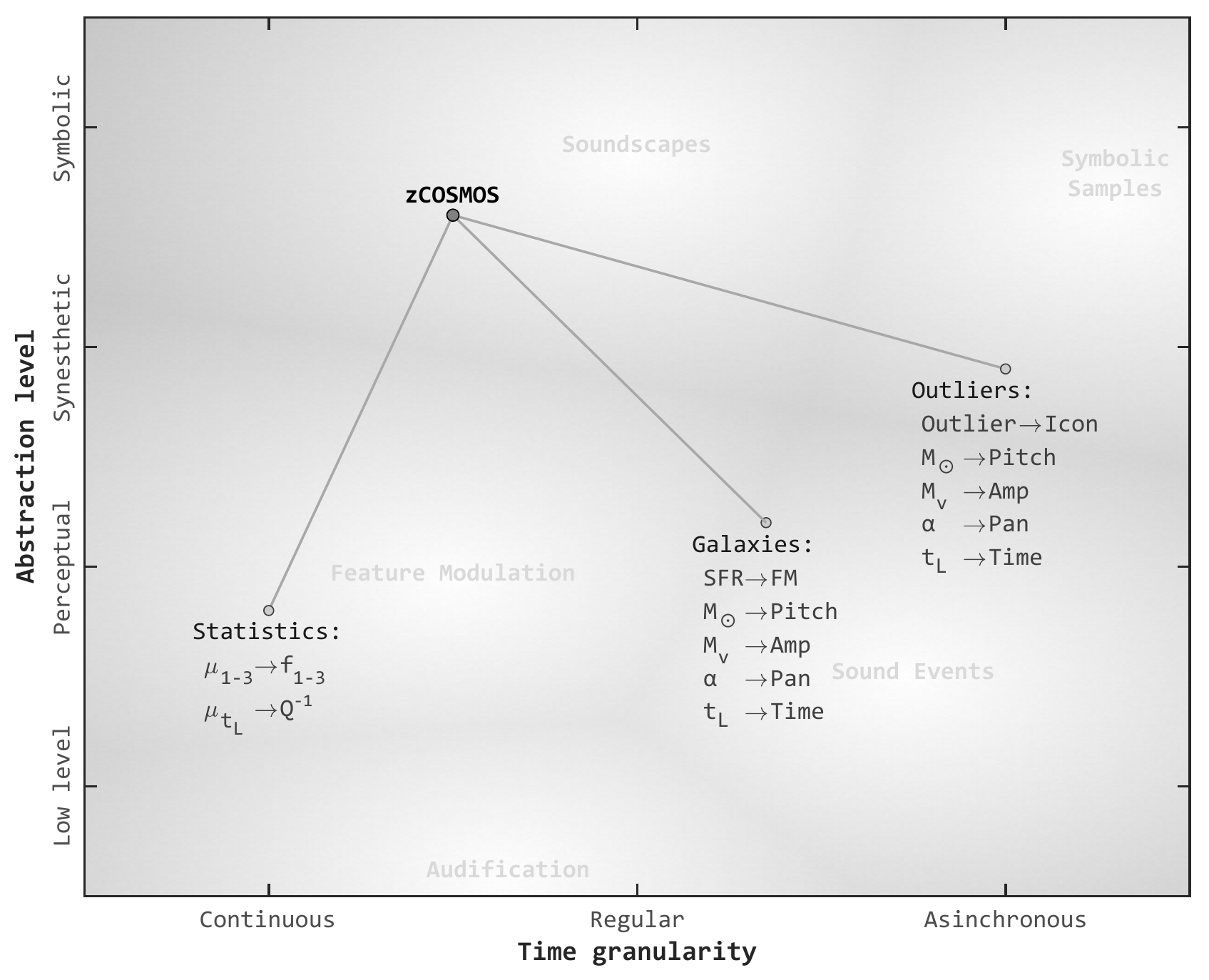}
	\caption{Representation in the sonification space. Three groups of data bindings with different levels of temporal granularity combine into one main feature which provides a comprehensive description of the dataset.}
	\label{fig:sonspace}
\end{figure}

As qualitatively depicted in Figure \ref{fig:sonspace}, galaxies, statistics, and outliers lay in three different areas of the sonification space. Sounds of galaxies are the constitutive atomic particles of the sonification, very small sounds which are meant to be mostly heard in swarms. The sound driven by statistics represents the scene, the space where all takes place, bent by the cumulative properties of the galaxies. Finally, outliers events are infrequent distinctive sounds, counterpointing the swarm movements. 

The outcome, marked as \zCOSMOS\ in Figure \ref{fig:sonspace}, can be perceived as both an artificial soundscape and a contemporary music composition. 

This sonification is like a journey back in time, which starts from the present, close to the Earth, and proceeds away, revealing a more and more distant past.

From a musical point of view, an overall crescendo can be clearly perceived, with a number of intermediate full-orchestra moments. Such a crescendo effect is due to both the increasing in average dynamics linked to each event (see the dotted line in Figure \ref{fig:stats}) and to the increasing density of events (see the shape of the gray area in Figure \ref{fig:stats}). In this sense, a 25-minute-long listening session is a good compromise to make the general dynamic trend emerge without having an excessive density of sound events. Moreover, the dashed line in \ref{fig:stats} shows that also the average frequency-modulation amount increases with time, thus producing a darker sound at the beginning and a brighter and richer sound at the end of the piece.

The trend shown by the three lines in Figure \ref{fig:ft}, that correspond to the frequencies of drone sounds, creates a contrary-motion effect between the bass line and the leading voices of the background layer. 

In order to make the listening experience more interesting, the role played by the sonification of outliers is fundamental. In fact, in musical terms, the function of outliers can be compared to sudden variations in the orchestration, impulsive percussive events, and articulation signs. In particular, the outlier related to the biggest galaxy, rendered as a low frequency percussion, happens to be played only once, and very close to the end of the sonification: this sound event recalls the musical function of a final cadence.

\section{Conclusions}
\label{sec:conclusions}

In this work we have described the multiple stages bringing from the design to the realization of a sonification driven by astronomical data.

The whole process is articulated and requires heterogeneous competences. The working group embraced domain experts in different fields (physicists and astronomers, music composers, performers, sound and music computing experts, etc.) able to share ideas and cooperate. 

The main activities conducted to realize the sonification (and the key actors involved) have been:

\begin{itemize}
    \item the acquisition of the dataset (physicists and astronomers);
    \item its transformation in a suitable computer format (computer scientists);
    \item its filtering according to the scientific and aesthetic goals of the sonification (artists and sound-and-music computing experts);
    \item design choices concerning data bindings (the whole working group);
    \item data pre-processing (physicists, astronomers and computer scientists);
    \item technical production of single sounds (sound designers);
    \item mixing and post-production (sound-and-music computing experts).
\end{itemize}

This project has been mainly conceived for scientific dissemination. The goal of a sonification activity in general, and of the \zCOSMOS\ initiative in particular, is to make scientific data, originally hard to retrieve and understand, easily accessible and enjoyable even by a non-expert audience. Public presentations are planned to take place in cultural institutions, together with explanatory wall-mounted panels (for museum-like installations) or support videos (for planetariums).

Thanks to the choice to diverge from tonal-harmony rules, rather focusing on the best possible rendering of the original data, this sonification challenges the listener by proposing uncommon musical structures typical of contemporary music. Such a listening activity is expected to raise users' awareness about their own perception, and highlight the importance of sound as a carrier of meaning.


\bibliographystyle{plain}
\bibliography{references}
\end{document}